# On-Surface Synthesis of Heptacene on Ag(001) from Brominated and Non-Brominated Tetrahydroheptacene Precursors

Luciano Colazzo,[a,b] Mohammed S. G. Mohammed,[a,b] Ruth Dorel,[c] Pawel Nita,[a,b] Carlos García Fernández,[a] Paula Abufager,[d] Nicolás Lorente,[a,b] Antonio Echavarren,[c,e] and Dimas G. de Oteyza*[a,b,f]

[a] *Donostia International Physics Center (DIPC), 20018 San Sebastián, Spain.*
[b] *Centro de Física de Materiales (CSIC-UPV/EHU) - MPC, 20018 San Sebastián, Spain.*
[c] *Institute of Chemical Research of Catalonia (ICIQ), Barcelona Institute of Science and Technology, 43007 Tarragona, Spain.*
[d] *Instituto de Física de Rosario, Consejo Nacional de Investigaciones Científicas y Técnicas (CONICET) and Universidad Nacional de Rosario, (2000) Rosario, Argentina*
[e] *Departament de Química Orgànica i Analítica, Universitat Rovira i Virgili, 43007 Tarragona, Spain*
[f] *Ikerbasque, Basque Foundation for Science, 48011 Bilbao, Spain.*

**Achieving the Ag(001)-supported synthesis of heptacene from two related reactants reveals the effect of the presence of Br atoms on the reaction process. The properties of reactant, intermediates and end-product are further characterized by scanning tunneling microscopy and spectroscopy.**

Acenes are a class of polycyclic aromatic hydrocarbons consisting of linearly fused benzene rings. Their interesting electronic properties have made them subject of a vast number of experimental and theoretical studies.[1] By way of example, pentacene is among the best studied organic semiconductors and has been successfully integrated into several types of devices, outstanding for its particularly high charge carrier mobility.[1] However, in spite of their increasingly promising electronic properties predicted by theoretical calculations,[2] acenes longer than pentacene have been studied much less. This is directly related to their lower stability,[3] which makes their synthesis by conventional wet chemistry much more challenging,[4] and further limits their potential implementation into device structures. Indeed, the synthesis of non-substituted higher acenes has been until recently restricted to the photodecarbonylation of carbonyl-bridged precursors, which needs to be carried out in stabilizing inert matrices at cryogenic temperatures.[5] Among the reasons for such reduced stability is the increasing open shell character predicted for acenes as they grow longer. According to calculations, partial diradical character starts appearing for hexacene, with a notable contribution of polyradical character for longer acenes like undecacene or dodecacene.[6] That is, predictions associate roughly two unpaired electrons every five to six rings.[7]

On-surface chemistry under ultra-high vacuum (UHV) has appeared as an efficient way to overcome some of those limitations. Under the clean and controlled environment of UHV chambers, higher acenes have not only been successfully synthesized, but are also stable, allowing their subsequent characterization with surface science techniques. Following that approach, tetracene,[8] hexacene,[9] heptacene,[10,11] octacene,[11] nonacene,[11,12] decacene[11,13] and undecacene[11] have been synthesized and characterized on Au(111). Pentacene and heptacene have also been synthesized on Ni(111)[14] and Ag(111),[15] respectively.

In this work, starting from two different precursors, we study the synthesis process of heptacene on Ag(001). Single molecule analysis by scanning tunnelling microscopy (STM) and spectroscopy (STS) has been applied to extract information on the structural and electronic properties of reactant, intermediates, as well as of the heptacene end-product.

The precursors correspond to dibromotetrahydroheptacene (Br-**1**, Fig. 1a) and tetrahydroheptacene (**1**, Fig. 1d),[16] merely differing in the bromination (or its absence) at the central carbon ring. Deposition of the reactants on a surface held at room temperature (RT) and subsequently cooled to 4.3 K for its characterization results in samples as shown in Fig. 1. The samples with Br-**1** (Fig. 1b,c) display two types of well-differentiated surface morphologies: reconstructed substrate regions displaying highly periodic monatomic steps along the compact [110] or [1-10] directions (see also Fig. S1) and flat Ag(001) terraces. Unless high coverages force the formation of close-packed layers all over the substrate, the molecules appear preferentially adsorbed on the reconstructed regions. The molecules all appear readily debrominated at RT. While this is similar to the findings with other brominated reactants on Ag(110),[17] it differs from results on Ag(111),[17,18] suggesting, as generally expected, a higher reactivity of less dense-packed surfaces. Each reconstructed terrace fits one molecule and one to three neighbouring Br atoms. Molecules on subsequent terraces appear aligned with respect to one another and display long-range order domains.

Equivalent experiments with the non-halogenated reactant **1** lead to notably different results (Fig. 1d,e). In the absence of Br substituents, the molecules appear on the Ag(001) terraces aligned along the [110] or [1-10] directions, but cause neither a surface reconstruction nor display any kind of long-range order. We thus associate these two effects to Br-driven interactions, in line with previous reports of other brominated precursors on nobel metal surfaces.[19]

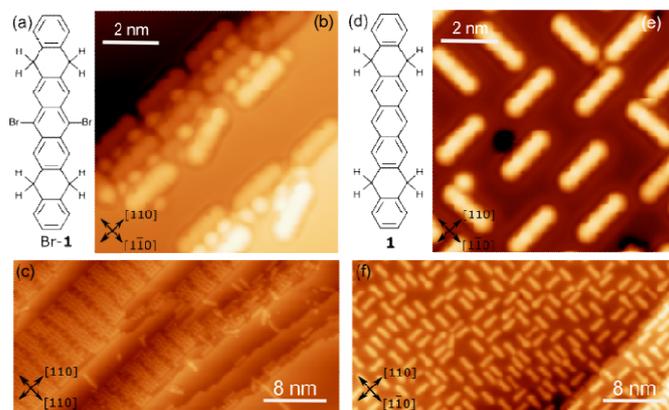

**Fig. 1**. (a) Chemical structure of the precursor Br-**1** and representative small (b) and large (c) scale STM images of the surface after its deposition on Ag(001) held at RT [U=30 mV / I=70 pA (b) and U=30 mV / I=100 pA (c)]. (d) Chemical structure of the reactant **1** and representative small (e) and large (f) scale STM images of the sample after its deposition on Ag(001) [U=35 mV / I=100 pA (e) and U=600 mV / I=600 pA (f)].

When heated, both kinds of reactants suffer dehydrogenation of the non-aromatic rings. In this process, hydrogens are lost pairwise, while on the organic scaffold the aromatization of the rings occurs as displayed in Fig. 2a. Focusing first on Br-**1**, a stepwise dehydrogenation is observed for the two rings. Although a detailed characterization of the threshold temperatures was not performed, annealing of the samples to 180ºC caused the aromatization of one ring in most of the molecules (Fig. 3a), and a quasi-full transformation into heptacene was observed after annealing to 270ºC. The disparate activation temperatures evidence differences in the mechanistic of the first and second aromatization events, which may be affected e.g. by potentially remaining radicals generated upon dehalogenation. After aromatization of the first ring, the remaining non-aromatic ring easily shifts along the molecule (Fig. 2h-i). Its position can be distinguished through high-resolution constant height imaging using CO-functionalized tips, which features the methylene functionalities on the polyacenes as distinctly wider rings (Fig. 2a,b). Alternatively, an indentation-like contrast at low bias (highlighted with arrows in Figs. 2d-i), an increased apparent height at positive or a node at negative bias (Fig. S2) is observed at the hydrogenated rings by conventional constant current STM imaging. A statistical analysis of the position of the hydrogenated ring reveals its tendency to shift toward the central part of the molecule (Fig. 3a,b). Since the hydrogenation breaks the conjugation along the molecule, intermediates **2-4** can be seen as made up by two coupled but independent acenes. Taking into account that the aromatic stabilization energy per π-electron as calculated from homodesmic reactions is notably reduced as the number of annulated rings increases,[20] the shorter the acene segments to be combined are, the more stable the molecules are. As a result, occurrence is highest for **4**, followed by **3** and then by **2**, whose longest acene segments correspond to anthracene, tetracene and pentacene, respectively. In fact, even those few molecules that retain the tetrahydrogenation after this first annealing step (e.g. oval in Fig. 3a) display the hydrogenation preferentially at rings 2 and 3 on either side of the molecular center (intermediate **1´**, Fig. 2g), reducing the maximum conjugation length from an anthracene to a naphthalene segment. If heated to higher temperatures (e.g. 270ºC), the remaining non-aromatic ring also gets dehydrogenated, rendering heptacene as the end-product **5** (Fig. 2c,f).

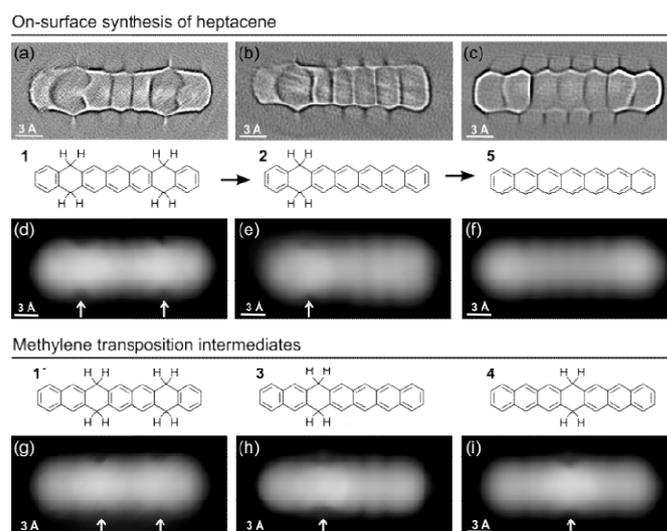

**Fig. 2**. On-surface synthesis process of heptacene from Br-**1**, displaying the evolution of the molecule´s chemical structure throughout the reaction with two, one, and no hydrogenated non-aromatic rings. The associated high resolution, constant height images revealing the bonding structure (U=2 mV, Lapplace filtered) are shown in panels (a), (b) and (c), respectively. Panels (d), (e) and (f) display the corresponding constant current STM images (U=30-50 mV / I=100 pA), while (g), (h) and (i) display the intermediates involving methylene transposition. All images are recorded with CO-functionalized probes.

Interestingly, the scenario changes substantially in the case of "kinked acene isomers" (Fig. 3c). Analyzing a sample containing minor amounts of tetrahydroheptaphene (see Suppl. Infor. for details) we observed that, in contrast to the linear counterparts, the saturation of the dihydro intermediates always remains at ring 2 (Fig. 3b). In this case, any hydrogen migration would reduce the number of Clar sextets in the molecule from three to two (Fig. 3d), corroborating that the stabilization energy per Clar sextet[20,21] makes the configuration with the methylenes on ring 2 most favorable.

Also the non-halogenated precursor **1**, still following a similar reaction process as Br-**1**, displays notable differences. As opposed to the Br-**1** case displayed in Fig. 3a, a similar annealing to 180ºC results in an approximately 80/13/7 ratio of tetrahydro, dihidro and heptacene species (Fig. S3). That is, at the same temperature, less molecules display any aromatization, and out of those that do, more aromatize fully into **5**. Besides, the formal methylene migration on reactant and dihydro-intermediates is much less common when starting from **1**. Thus, although the same tendency to minimize the conjugated acene segments through formal methylene migration was reported for closely related non-brominated tetrahydrononacene molecules on Au(111),[12] and although an analysis of the detailed mechanism involved in these chemical changes is beyond the scope of this paper, it seems evident that the initial presence of radicals after the dehalogenation facilitates the first aromatization and lowers the hydrogen transposition barriers.

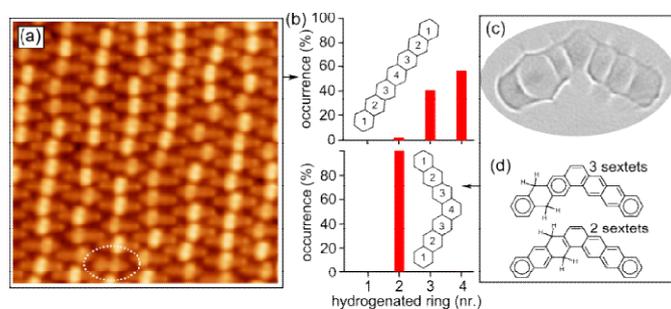

Fig. 3. (a) STM image (U=1.7 V / I=186 pA) of a full monolayer sample of Br-**1** after annealing to 180ºC. (b) Histograms displaying the occurrence of differently positioned hydrogenated rings (numbered as in the molecular structure insets) for dihydro intermediates of heptacene (top) and heptaphene (bottom). (c) Laplace filtered constant-height STM image (U=2 mV) of a heptaphene intermediate with a single hydrogenated ring. (d) Schematics revealing the maximized number of Clar sextets for dihydroheptaphene when hydrogenated at the second ring (compared below, e.g., with hydrogenation at the third ring).

Within these experiments, we end up with a plethora of different but closely related molecules, offering a good opportunity to probe structure-property relations. At this stage it is important to note that, although with the slight differences in the reaction process described above, the structural and electronic characterization performed on single molecules on Ag(001) terraces is similar regardless of whether **1** or Br-**1** are used. That is, we observe products from a spontaneous hydrogenation of the radical sites resulting from the dehalogenation of Br-**1.** This is of key importance in the analysis and understanding of many surface-supported reactions under vacuum and may be rationalized taking into consideration that hydrogen makes up for most of the residual gas present in UHV chambers and may be additionally generated by nearby dehydrogenation events.

We have recorded dI/dV STS spectra on each of those different molecules for comparison. The results are summarized in Fig. 4a, placed in order according to the molecule´s longest conjugated segment, which is known to host most of the electron density of the highest occupied (HOMO) and lowest unoccupied molecular orbitals (LUMO) (Fig. S4).[12] Thus, we start from **1´**, displaying naphthalene as the largest conjugated segment, followed by **1** and **4** (anthracene), **3** (tetracene), **2** (pentacene) and **5** (heptacene). For the sake of comparison, we also add heptaphene (**6**), the fully conjugated kinked isomer of heptacene (**5**).

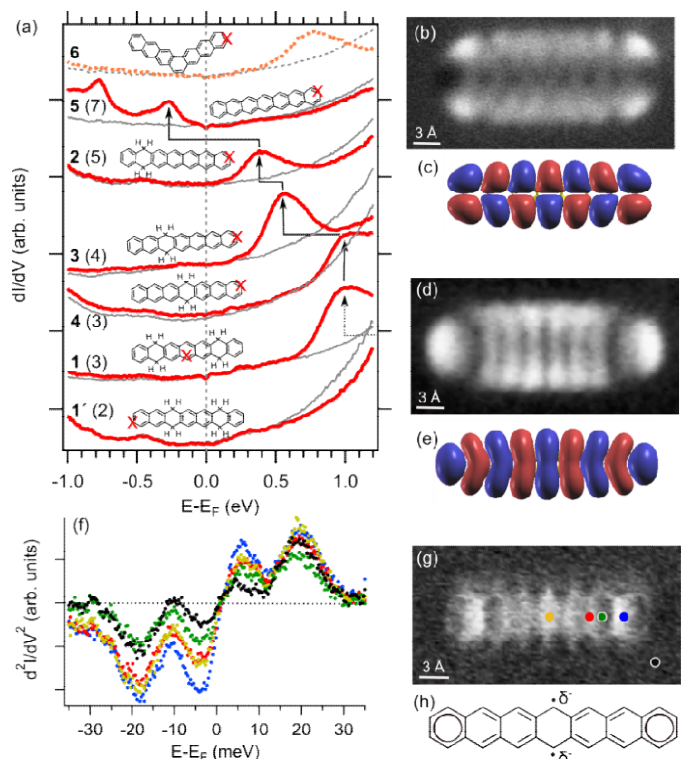

**Fig. 4**. (a) dI/dV point spectra on reactant, intermediates and end-product, plus heptaphene **6** (colored lines) at positions as marked on the accompanying models, and associated reference substrate spectra (grey lines). The nr. of rings in the longest conjugated segment is given in brakets and the LUMO position is marked with arrows. Conductance maps (I=180 pA, $U_{osc}$=10 mV, ⍵=341 Hz) of **5** at -0.76 V (b) as well as at -0.25 V (d) and its calculated gas-phase wavefunction of HOMO (c) and LUMO (e). (g) $d^2I/dV^2$ spectra ($U_{osc}$=2 mV, ⍵=341 Hz) at the positions marked in (g), revealing the frustrated translational and rotational modes of the CO on the functionalized tip at around 3 and 20 meV, respectively. (g) Inelastic electron tunneling spectroscopy map (const. z, $U_{osc}$=7 mV, ⍵=341 Hz) of **5** at 3.5 mV. (h) Schematic model of the open-shell heptacene structure, with the associated radicals partially stabilized by additional electron charge.

The HOMO-LUMO gap is known to decrease with the size of acenes,[5c,11,12] as reproduced also with density-functional theory calculations of these specific molecules (Fig. S4). Although experimentally we have only accessed the LUMO for most of the probed molecules, it can be nicely observed how it approaches the Fermi level as the largest conjugated segment increases in size from **1** to **2**.[22,23] Following the same trend, the shift from **2** to **5** occurs by crossing the Fermi level, concurrently to the appearance of a second occupied resonance. Conductance maps at the corresponding energies (Fig. 4b and d) reveal their resemblance with the calculated gas-phase wavefunctions of HOMO and LUMO (Fig. 4c and e), the slight differences in the representations stemming from the mixed s- and p-wave character of the CO-functionalized probe used for the characterization.[24] These data are thus clear evidence of the charging of heptacene on Ag(001) by filling of its former gas-phase LUMO.

Such important charge transfer suggests a strong molecule-substrate interaction,[25] as opposed to weakly interacting interfaces that rather display a Fermi level pinning scenario.[26] The strong interaction is proposed to be promoted by the partial open shell character of heptacene, whose radicals could be partially stabilized by the additional electrons as schematically shown in Fig. 4h. Experimental findings hinting at this particular scenario are the spatial variations in the vibrations of CO at functionalized tips when performing inelastic electron tunneling spectroscopy (IETS).[27] Such vibrations, and in particular the frustrated translational mode occurring at around 3 mV at the silver-CO-tip gap, is strongly dependent on the charge density of the probed sample and shows strongest intensity variations at the outer and central rings (Fig. 4g,h), where the Clar sextets and the partial radical character are expected to be dominantly located.

Altogether, we have studied the surface-supported synthesis of heptacene on Ag(001). The use of two different precursors has allowed us to learn about the effect of bromine atoms on the properties of molecule and substrate, as well as about the spontaneous hydrogenation of radical sites even under UHV. Spectroscopic analysis of the reactant, intermediates and end-product has evidenced a change in the energy level alignment associated with the decreasing band gap of molecules with increasing conjugation size. That trend ends with a charged heptacene, indicative of a strong interaction with the underlying Ag(001) substrate.

We acknowledge funding from the European Research Council under the European Union's Horizon 2020 programme (grant agreement No. 635919), from the Spanish Ministry of Economy, Industry and Competitiveness (MINECO, Grant Nos. MAT2016-78293-C6, MAT2015-66888-C3-2-R), Agencia Estatal de Investigación (AEI)/FEDER, UE (CTQ2016-75960-P), the AGAUR (2017 SGR 1257), and CERCA Program / Generalitat de Catalunya.

## Notes and references

# Supplementary Information:

# On-Surface Synthesis of Heptacene on Ag(001) from Brominated and Non-Brominated Tetrahydroheptacene Precursors.

Luciano Colazzo,[a,b] Mohammed S. G. Mohammed,[a,b] Ruth Dorel,[c] Pawel Nita,[a,b] Carlos García Fernández,[a] Paula Abufager,[d] Nicolás Lorente,[a,b] Antonio Echavarren,[c,e] and Dimas G. de Oteyza *[a,b,f]

[a] Donostia International Physics Center (DIPC), 20018 San Sebastián, Spain.

[b] Centro de Física de Materiales (CSIC-UPV/EHU) - MPC, 20018 San Sebastián, Spain.

[c] Institute of Chemical Research of Catalonia (ICIQ), Barcelona Institute of Science and Technology, 43007 Tarragona, Spain.

[d] Instituto de Física de Rosario, Consejo Nacional de Investigaciones Científicas y Técnicas (CONICET) and Universidad Nacional de Rosario, (2000) Rosario, Argentina

[e] Departament de Química Orgànica i Analítica, Universitat Rovira i Virgili, 43007 Tarragona, Spain

[f] Ikerbasque, Basque Foundation for Science, 48011 Bilbao, Spain.

**Precursor synthesis**

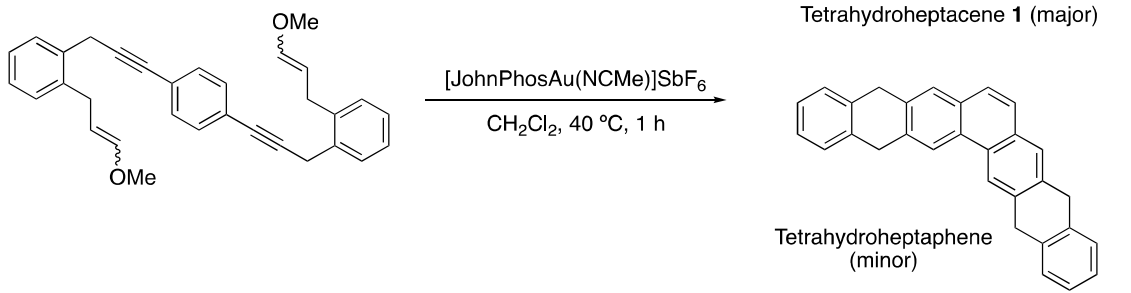

Scheme S1. The sample containing a mixture of tetrahydroheptacene (**1**) and kinked tetrahydroheptaphene was obtained by following the procedure reported for the preparation of **1**[1] with a modified purification protocol. Thus, the crude mixture was washed and centrifuged with MeOH (2x5 mL) and a 4:1 *v/v* MeOH:CH$_2$Cl$_2$ mixture (5 mL). The remaining solid was subsequently dried under reduced pressure to afford a white solid that consists of **1** as the major product together with minor amounts of kinked tetrahydroheptaphene.

**Determination of the position of hydrogenated rings from conventional STM imaging.**

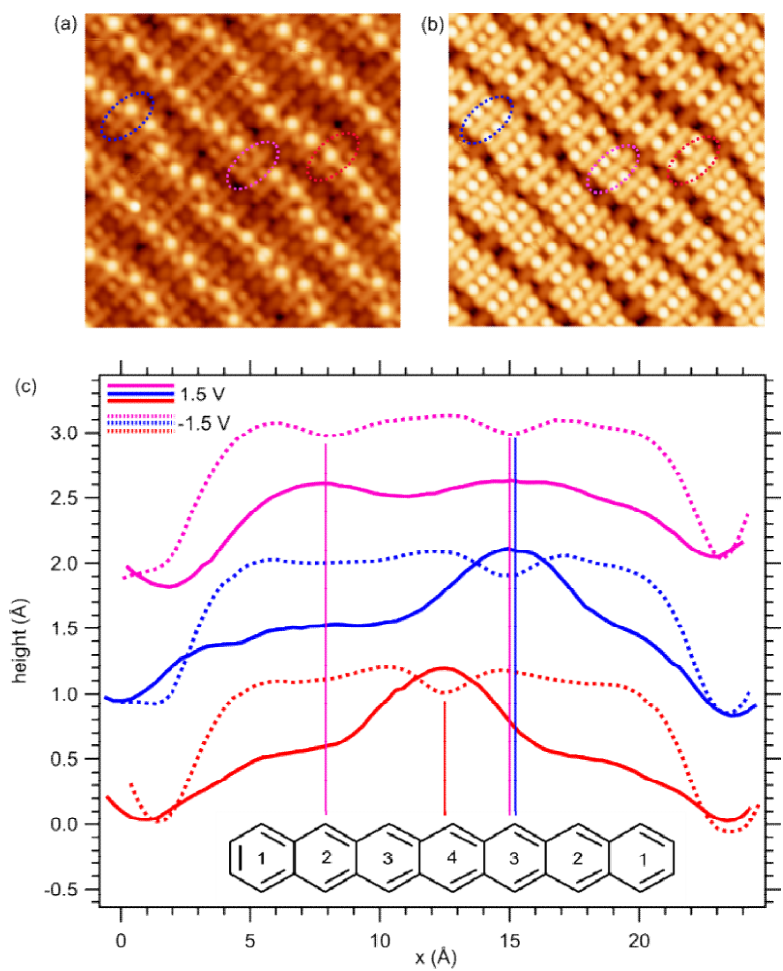

**Figure S1**. Conventional constant current STM imaging of heptacene intermediates at (a) U = 1.5 V and (b) U = -1.5 V sample bias. The former displays protrusions at the hydrogenated ring, while the latter displays a node. Profiles on the surrounded molecules in (a) and (b) are further displayed in (c), put in relation with a molecular structure model. It can be seen how the same profiles on the same molecules display lobes or nodes at positive or negative bias, respectively, and the correspondence with the particular rings.

# Surface morphology variations: comparison between precursors Br-1 and 1 on Ag(001)

The surface morphology resulting from the deposition of precursor Br-1 reveal substantial differences when compared with the one resulting after the deposition of precursor 1. In Figure S1 representative STM images are shown for the deposition of precursor Br-1 (top-panel) and precursor 1 (bottom-panel) on Ag(001) held at RT, for two different coverages: below the 50% of the full terrace area coverage ($\Theta$) (a,c left-panel) and close or above the $\Theta$ = 50% (b,d right-panel). The trivial difference in the two cases being that the molecules of precursor Br-1 result massively confined in step regions, while precursor 1 predominantly adsorbs on flat regions of the Ag(001) surface.

In Figure S1a an example of surface morphology variation it is shown following the deposition of precursor Br-1 aiming at a coverage $\Theta \leq 10\%$. The flat terraces of Ag(001) remains clean and the molecules are generally found confined in the regions highlighted by the black arrows. A closer analysis of these regions reveals the formation of mono-atomic steps with high density on the Ag(001) surface (see main text Figure 1). By increasing the surface coverage (Figure S1b, $\Theta \geq 50\%$), the extension of the clean flat terraces of Ag(001) is reduced, still, it remain unpopulated by molecules and the extension of the stepped region increases (black arrows highlight a wider region).

The system is notably different when compared with the one resulting from the deposition of precursor 1, a s shown in Figure S1c, since for a comparable coverage as the one of Figure S1a, the molecules are mainly found on the flat terraces of the Ag(001) surface and only decorate the monoatomic step edges. In this system the surface morphology of the Ag the surface morphology is not affected even for higher molecular coverages as show in Figure S1d.

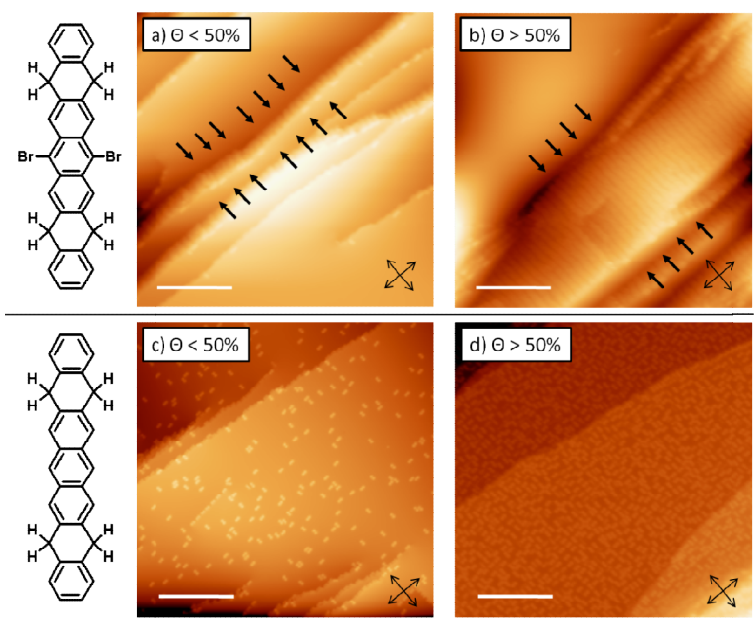

**Figure S2**: comparison of the surface morphology variations resulting from the deposition of precursor **Br-1** (top) and precursor **1** (bottom) on Ag(001) held at RT, for two different coverages: low (left) and high (right). (a),(b),(c),(d) STM images [scale bar 20nm; all frames recorded at U=1000 mV / I=100pA; 80×80 nm$^2$]. Square cross arrows represent the high symmetry [110] and [1-10] directions of the substrate. In (a) and (b) black arrows are a guide for the eye used to highlight the extension of the high density step regions at increasing coverages.

**Sample after annealing reactant 1 to 180ºC**

As opposed to the findings when starting from Br-**1**, displayed in Fig. 3 of the main paper, annealing **1** to 180ºC results in a mixture of tetrahydro, dihydro and fully aromatized molecules in a ratio around 80/13/7. This is exemplified in Fig. S3.

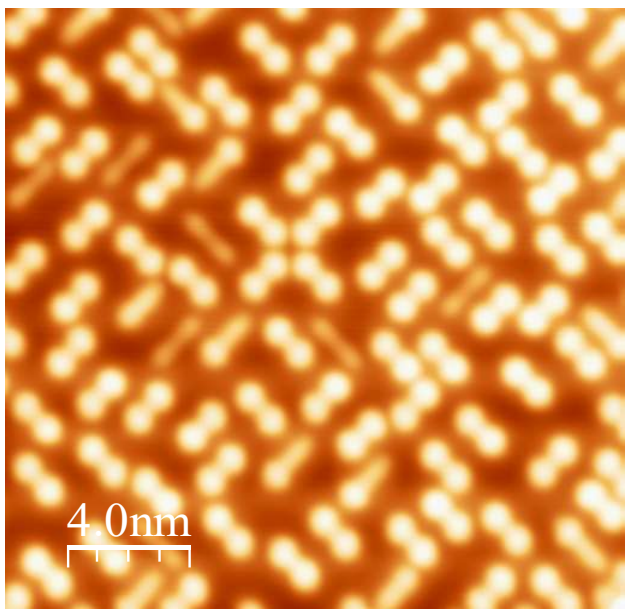

**Figure S3**: Constant current STM image of a Ag(001) sample after deposition of 1 and subsequent annealing to 180ºC. The tetrahydro, dihydro and fully aromatized molecular species can be discerned from their two bright lobes, one bright lobe, or no lobes together with a lower apparent height, respectively.

**Details on calculations**

The theoretical electronic structure of gas-phase reactant, intermediates and heptacene molecules have been obtained with the plane-wave package QUANTUM ESPRESSO2[2] using the generalized gradient approximation (GGA) to the exchange-correlation functional in the parametrization of Perdew, Burke, and Ernzerhof (PBE).[3] Ultrasoft pseudopotential[4] with a plane-wave cutoff of 35 Ry for the wave functions and 350 Ry for the charge density were used. All atoms in the molecule were relaxed until forces were smaller than 0.01 eV/ Å.

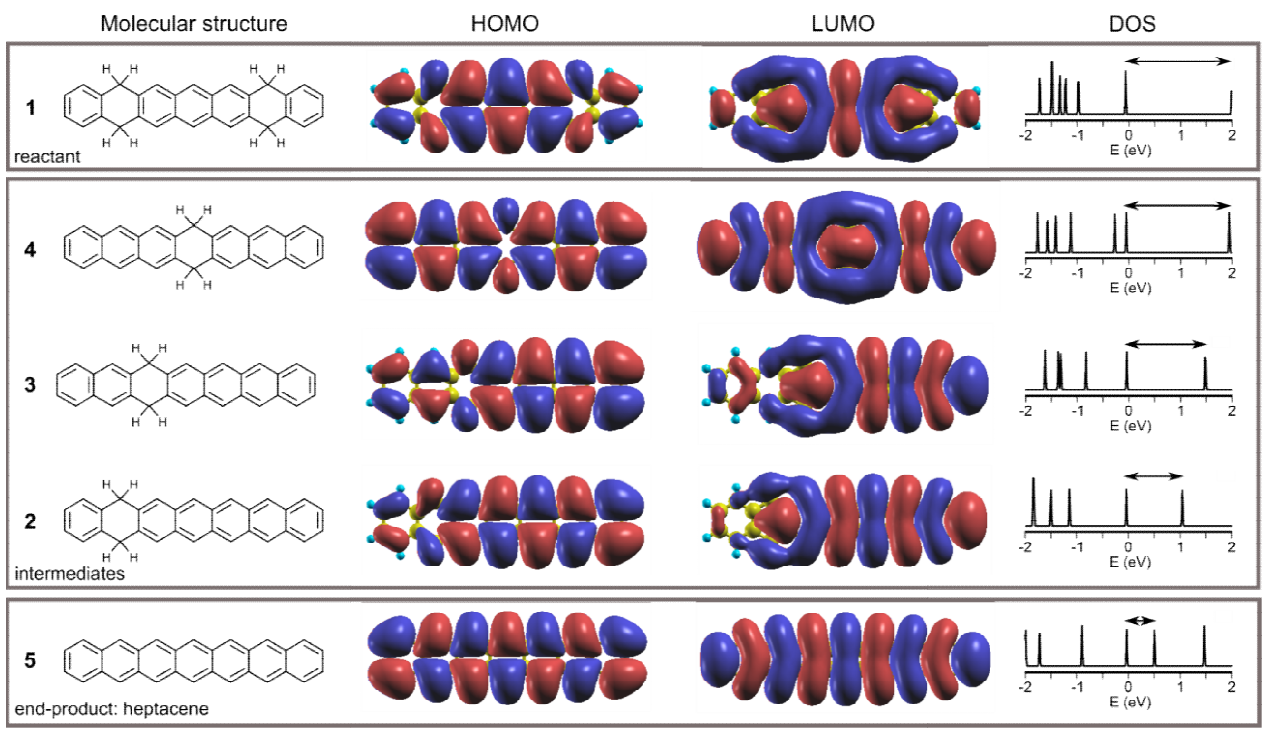

**Figure S4**. Molecular structure, calculated wavefunctions for gas-phase HOMO and LUMO of reactant, intermediates and product. The right-most column additionally displays the density of states of the respective molecules, with the HOMO level at 0 as reference, revealing how the LUMO shifts to higher energies (increasing the HOMO-LUMO gap, indicated by the arrows) as the longest conjugated segment (which is observed to host most of the HOMO and LUMO density of states) shortens. Interestingly, in line with the experimental observations (Fig. 4 of main text), the gap hardly changes from **1** to **4**, both featuring three rings as the longest conjugated segment.